# Chaos Engineering

Ali Basiri[1], Niosha Behnam[1], Ruud de Rooij[1], Lorin Hochstein[1], Luke Kosewski[1], Justin Reynolds[1], Casey Rosenthal[1]

[1]Traffic and Chaos Team, Netflix

## Abstract

Modern software-based services are implemented as distributed systems with complex behavior and failure modes. Many large tech organizations are using experimentation to verify the reliability of such systems. We use the term "Chaos Engineering" to refer to this approach, and discuss the underlying principles and how to use it to run experiments.

## Introduction

Thirty years ago, Jim Gray noted that "A way to improve availability is to install proven hardware and software, and then leave it alone" [1]. For companies that provide services over the Internet, "leaving it alone" isn't an option. Such service providers must continually make changes in order to increase the value of the service, such as adding new features and improving performance. At Netflix, engineers push new code into production and modify runtime configuration parameters hundreds of times a day. Availability is still important: a customer who can't watch a video because of a service outage might not be a customer for long. But to achieve high levels of availability, we need to apply a different approach than what Gray advocated.

For years, Netflix has been running an internal service called Chaos Monkey [2], which randomly selects virtual machine instances that host our production services and terminates them. Chaos Monkey's purpose was to encourage Netflix engineers to design software services that can withstand failures of individual instances. Chaos Monkey is only active during normal working hours so that engineers can respond quickly if a service fails due to an instance termination.

Chaos Monkey proved successful, as today inside of Netflix all engineers design their services to handle instance failures as a matter of course. That success encouraged us to extend the approach of injecting failures into the production system in order to improve reliability. For example, we perform "Chaos Kong" exercises that simulate the failure of an entire Amazon EC2 region, and we also run Failure Injection Testing (FIT) exercises where we cause requests between Netflix services to fail and verify that the system degrades gracefully.

Over time, we came to realize that there are common themes underlying these activities more subtle than simply "*break things in production*". We also noticed that organizations such as Amazon [3], Google [3], Microsoft [4], and Facebook [5], were applying similar techniques to test the resilience of their own systems. We believe that these activities form part of a discipline that is emerging in our industry, that we call **Chaos Engineering**. Specifically, *Chaos Engineering is the discipline of experimenting on a distributed system in order to build confidence in its capability to withstand turbulent conditions in production.* These conditions could be anything from a hardware failure to an unexpected surge in the number of client requests to a malformed value appearing in a runtime configuration parameter. We have posted online an overview of the Principles of Chaos Engineering [8], which we elaborate on in this article.

## Sidebar: Netflix and its architecture

Netflix provides customers access to movies and television shows, which are delivered by streaming over the Internet to devices such as smart TVs, set-top boxes, smartphones, tablets, and desktop computers.

The Netflix user interface displays content that is tailored to the specific user. For example, the list of movies on the home screen is customized based on data such as what the user has watched previously. The UI includes features such as search, recommendations, ratings, user profiles, and bookmarks for resuming previously watched videos.

There are two main components to the Netflix infrastructure:

1. Open Connect
2. Control Plane

*Open Connect* is Netflix's content delivery network (CDN), which is a geographically distributed cache of video data. Servers with large storage capacity are located at sites around the world. When a Netflix user hits the "play" button, these servers stream video data to the user.

The *Control Plane* is a distributed set of services that implements all the functionality in the UI other than streaming the the video itself. All of the services in the Control Plane run inside of virtual machines in the Amazon Web Services (AWS) cloud, replicated across three geographic regions to reduce latency and improve availability. The Chaos Engineering examples in this paper refer to services in the Control Plane.

# Shifting to a system perspective

At the heart of Chaos Engineering is the premise that engineers should view a collection of services running in production as a *single system*, and that we can better understand the behavior of this system by injecting real-world inputs (e.g., transient network failures, surges in incoming requests, malformed data inputs) and observing what happens at the system boundary.

In a traditional approach to software engineering, a functional specification describes how a software system should behave, as depicted in Figure 1. If the system is composed of multiple programs, the functional specification also defines the behavior of each individual programs.

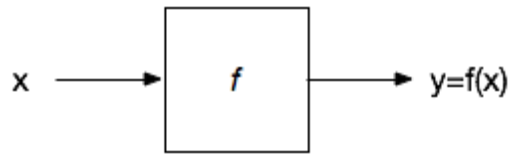

Fig. 1. A functional perspective of a software system. The specification *f* defines how inputs (x) map to outputs (y)

In practice, when describing distributed systems, functional specifications are incomplete; they do not fully characterize the behavior of the system for all possible inputs. The problem of incomplete specification is exacerbated by the scale and complexity of the user behavior and the underlying infrastructure, which make it difficult to fully enumerate all inputs. In practice, the scale of the systems makes it impractical to compose specifications of the constituent services in order to reason about behavior.

From a *system perspective*, engineers focus on the dynamic view of the software as a set of processes running in production. The system perspective assumes that the organization deployed the software into production some time ago, and that the system currently has active users. From this perspective the engineer views the system as a single entity and observes the behavior of the system by capturing metrics at the system boundary. The engineer cares about the typical behavior of metrics over time: the **steady-state behavior** of the system. The behavior can be changed by user interactions (e.g., rate of requests, type of requests), changes initiated by the engineers (e.g., runtime configuration changes, code changes), as well as other events such as the transient failure of a third-party service that the system depends on.

From this perspective, we ask questions like "does everything seem to be working properly?" and "are users complaining?" rather than "does the implementation match the specification?"

# Principles of Chaos Engineering

Chaos Engineering revolves around running experiments. As in other experimental disciplines, to design an experiment requires specifying [7]:

- hypotheses
- independent variables
- dependent variables
- context

We have identified the four principles which we believe embody the Chaos Engineering approach to designing experiments:

1. Build a hypothesis around steady state behavior
2. Vary real-world events
3. Run experiments in production
4. Automate experiments to run continuously

## Build a hypothesis around steady state behavior

At Netflix, we're using Chaos Engineering to ensure that the system still works properly under a variety of different conditions. However, "works properly" is too vague as a basis for designing experiments. In our context, the quality attribute we focus on most is *availability*.Netflix supports multiple features, including user interface layout customized to a particular user, recommendations about what content to watch, bookmarking that remembers where the user left off when watching previously, and a ratings system that shows how other users rated the show. Each of these features is implemented by a different service (or, in some cases, multiple services),  each of which can potentially fail. Yet, even if one of these internal services fails, this does not necessarily have an impact on overall system availability.

Netflix services use *fallbacks* to ensure graceful degradation: that failures in non-critical services have a minimal impact on the user experience. For example, consider a service *A* which makes requests against a caching service that sits in front of service *B*. if there is a failure in the caching service, then service *A* can fall back to making a request directly against service *B*. This behavior is not desirable in the long-term because it increases request latency and the load service *B*, but in the short term the user is not affected. In other cases, many Netflix services provide personalized content (e.g., personalized recommendations, personalized ratings), where the fallback would be to present a reasonable default. An example is the bookmark service, which allows users to resume watching a video at the location in the video where they previously stopped. If this bookmark service fails, then the Netflix user interface can default to starting videos at the beginning rather than providing a "resume from previous location" option.

Ultimately, what we care about is: "Are users able to find some content to watch, and successfully watch that content?" We operationalize this concept by observing how many users start streaming a video each second. We call this metric **SPS**, for (stream) starts per second [9].

We use SPS as our primary indicator of the overall health of the system. While the term "chaos" evokes a sense of unpredictability, one of the fundamental assumptions of Chaos Engineering is that complex systems exhibit behaviors that are regular enough that they can be predicted. Similarly, the SPS metric varies slowly and predictably over the course of a day, as shown in Figure 2. Engineers at Netflix spend so much time looking at SPS that they have developed an intuition about whether a given fluctuation is within the standard range of variation or whether it is cause for concern. If engineers observe an unexpected change in SPS, we know there is a problem with the system.

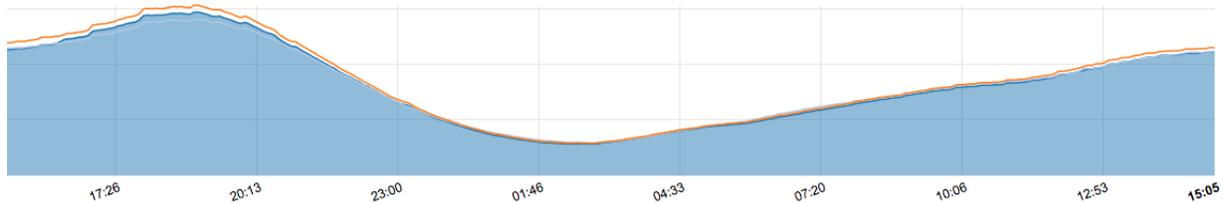

Fig. 2. A graph of SPS over a 24-hour period. The orange line shows the trend for the prior week. The y-axis is not labelled because of the proprietary nature of the data.

SPS is a good example of a metric that characterizes the **steady-state behavior** of the system. Another example of such a metric at Netflix is new account signups per second. There is a strong, obvious link between these metrics and the availability of the system: if the system is not available, users will not be able to stream video or sign up for the service. Other domains will use different metrics that characterize the system's steady-state behavior. For example, an e-commerce site might use a metric such as number of completed purchases per second, where an ad-serving service might use numbers of ads viewed by users per second.

When designing Chaos Engineering experiments, we form hypotheses around how the treatment will affect the steady state of the system. For example, Netflix has software deployments running in multiple geographic regions (Northern Virginia, Oregon and Ireland). If there is an outage that is confined to one of the geographical regions, we have the ability to fail over to another region: to redirect incoming client requests from the unhealthy region to one of the healthy regions. When we test this kind of failover scenario, we hypothesize that failing over from one region to another will have minimal impact on SPS.

When we formulate a hypothesis for a Chaos Engineering experiment, the hypothesis is about a very particular kind of metric. Compare a metric like SPS that characterizes the steady-state of the overall system with a finer-grain metric such as CPU load or time to complete a database query. Chaos Engineering experimental designs are focused on the former: steady state characterizations that are visible at the boundary of the system, which directly capture an interaction between the users and the system. We don't use these type of finer grained metrics for the design of Chaos Engineering experiments, since they are not a direct measure of system availability. However, we do observe finer-grained metrics when running Chaos Engineering experiments to check if the system is functioning properly, since these metrics indicate impact at the service level that isn't felt by the user. For example, a an increase in request latency or CPU utilization may be symptoms that a service is operating in degraded mode, even though from the user's point of view, the system is working properly. Individual service owners at Netflix set up alerts on these internal metrics to catch problems with their particular services. We may even conclude an experiment early if fine-grained metrics indicate that the system is not functioning correctly, even though SPS has not been impacted.

## Vary real-world events

The "happy path" is a term of art in software development that refers to an execution trace through a program where the inputs do not lead to any errors or corner cases. When engineers test their own code, the temptation is very great to focus their testing effort on the happy path. This leads to code that works for common cases.

Unfortunately, those corner cases and error conditions will happen out in the real world, even if they don't happen in our test cases. Clients of our services send malformed requests, services that we consume send malformed responses, the servers we run on die, or their hard disks fill up, or their memory is exhausted, network latencies temporarily spike by several orders of magnitude, and traffic from our clients can spike unexpectedly. A recent study reported that 92% of catastrophic system failures were the result of incorrect handling of non-fatal errors [10].

This brings us to the next chaos principle: vary real-world events. When designing Chaos experiments, choose your stimulus by sampling from the space of all possible inputs that might occur in the real world. One obvious choice is to look at historical data to see what type of inputs were involved in previous system outages, to ensure that the kinds of problems that happened previously can't happen again. Providing access to this historical data is yet another reason for doing post-mortems on system outages.

However, any input which you think could potentially disrupt the steady-state behavior of the system is a good candidate for an experiment.

Some examples of inputs we use at Netflix in our experiments:
- terminate virtual machine instances
- inject latency into requests between services
- fail requests between services
- fail an internal service

- make an entire Amazon region unavailable

Note that while our experiments at Netflix to date have focused on hardware and software failures as inputs, we believe that using other kind of inputs is valuable as well. For example, we could imagine other types of inputs such as the rate of requests received by a service, making changes to runtime parameters, or making changes to metadata that propagates through the system.

In some cases, you may need to simulate the event instead of inject it. For example, at Netflix we do not actually take an entire Amazon region offline, since we don't have the capability to do so. Instead, we take actions under the assumption that a region has gone offline: redirecting client requests to other Amazon regions, and we observe the effects. Ultimately, the engineers designing the chaos experiments must use their judgment to make trade-offs between the realism of the events and the potential risk of harm to the system. An interesting illustration of this tradeoff is the work done by Parrish and Halsey to run this type of experiment on a production financial trading system [11].

In other cases, we selectively apply the event to a subset of users. For example, as part of an experiment we may make an internal Netflix service behave as if it is unavailable from the point of view of some users, but for others the service appears to function properly. This allows us to reduce the scope of the experiment to a subset of users as a risk mitigation strategy.

## Run experiments in production

The computing power of a single server has increased dramatically in the past several decades. Despite these increases, it is not possible to deploy the entire Netflix system on a single server. No individual server has the compute power, memory, storage, network bandwidth, or reliability to host the Netflix system, which has millions of subscribers. The only known solution for implementing an Internet-scale service is by deploying software across multiple servers and having them coordinate over the network, resulting in a distributed system. With the rise in popularity of microservice architectures [12], we believe that more and more software engineers will be implementing Internet-based services as distributed systems.

Alas, traditional software testing approaches are not sufficient for identifying the potential failures of distributed systems. Consider the following examples of failure modes, based on observed failures inside of Netflix. In both cases, "client" and "server" refer to internal services, which typically act as both clients and servers.

1. A server is overloaded, and takes an increasing amount of time to respond to requests. One of the clients places outbound requests on an unbounded local queue. Over time, the queue consumes more and more memory, causing the client to fail.

2. A client makes a request to a service that is fronted by a cache. The service returns a transient error which is incorrectly cached. When other clients make the same request, they are served an error response from the cache.

These are examples of failure modes that require integration testing because they involve interactions among services. In some contexts, it may only be possible to do full integration testing in production. At Netflix, it is simply not possible to fully reproduce the entire architecture and run an end-to-end test.

However, when it is possible to reproduce the entire system in a test context, we still believe in the need to **run experiments in production**, since it is never possible to fully reproduce all aspects of the system within a test context. There will always be differences such as how synthetic clients behave compared to real clients, or DNS configuration issues.

### Automate experiments to run continuously

The fourth and final principle of Chaos Engineering is to leverage automation in order to maintain confidence in results over time.

Our system at Netflix changes continuously over time. Engineers modify the behavior of existing services, add new services, and change runtime configuration parameters. In addition, new metadata continually flows through the system as Netflix's video catalog changes over time. Any one of these changes can potentially contribute to a service interruption.

Because of these changes, our confidence in the results of past experiments decreases over time. While we expect that a new experiment will first be run manually, for the results of the experiments to be useful over time, the experiments will ultimately need to be automated in order to ensure that they run over and over again as the system evolves. The rate at which these experiments run will depend on context. For example, at Netflix, while we have Chaos Monkey running continuously during weekdays, we run Chaos Kong exercises at a cadence of once a month.

The idea of automating a Chaos experiment can be intimidating, since you are introducing some kind of stimulus which could lead to a system failure. However, Netflix's experience with Chaos Monkey suggest that this type of automation is feasible, and keeping it running ensures that new production services continue to be designed to withstand these types of failures.

## Running a Chaos experiment

Fundamental to Chaos Engineering is the execution of experiments to build confidence in system behavior.

Based on the principles, define an experiment:
1. Start by defining 'steady state' as some measurable output of a system that indicates normal behavior.
2. Hypothesize that this steady state will continue in both the control group and the experimental group.
3. Introduce variables that reflect real world events like servers that crash, hard drives that malfunction, network connections that are severed.
4. Try to disprove the hypothesis by looking for a difference in steady state between the control group and the experimental group.

As a hypothetical example, consider a service at Netflix that provides some value to the end-user but is not critical for video streaming, such as the bookmarks service that tracks the previous location where the user previously left off watching. We might design an experiment to verify that this service's failure will not have a significant impact on streaming.

For this experiment, we would use SPS as the measurable output and check that a failure in the bookmark service had only a minor impact on SPS. We identify a small percentage of Netflix users that will be involved in this experiment. We divide up the users into two groups, the control group and the experimental group. For the users in the experimental group, we will simulate the failure of the bookmark service by selectively failing all requests against the bookmark service associated with users in the experimental group. We can selectively fail requests for specific users through our failure injection testing service [6]. For the users in the control group, we do not introduce any failures. We hypothesize that the SPS values will be approximately equal for these two groups. We run the experiment for some period of time and then compare the SPS values.

At the conclusion of an experiment, we either have more confidence in the system's ability to maintain behavior in the presence of our variables, or a weakness has been uncovered and suggests a path for improvement.

# The Future of Chaos Engineering

While the concepts in this paper are not fundamentally new, the practice of applying these concepts in order to improve software systems is still in its infancy. This article is an attempt to explicitly define the underlying concepts in order to help advance the state-of-the-practice. We believe that the increasing complexity of software systems will continue to drive the need for empirical approaches to achieving system availability. Our hope is that the practitioner and research communities will come to recognize Chaos Engineering as its own discipline and to continue to move it forward, including:

**Case studies from other domains.** We know from informal discussions with other Internet-scale organizations that they are applying similar approaches. We hope that more organizations will document how they are applying Chaos Engineering to demonstrate that these techniques are not unique to Netflix.

**Adoption.** What approaches are successful for getting an organization to buy into this approach, and for getting engineering teams within the organization to adopt it?

**Tooling**. At Netflix, we are currently using tools built in-house which work with the infrastructure that we have built. It's not clear how much of the tooling of chaos experiments will be specific to a particular organization's infrastructure and how much will be reusable. Can we as a build a set of tools for running these types of experiments that are reusable across organizations?

**Event injection models**. The space of possible events to inject into a system is potentially quite large, especially when we consider combinations of events. Research shows that many failures are triggered by combinations of events rather than single events [10]. How do we decide what set of experiments to run?

If you would like to reach out to discuss Chaos Engineering, please contact us at chaos@netflix.com.

## References

[1] Jim Gray, Why do Computers Stop and What Can Be Done About It? Tandem Computers Technical Report 85.7, PN87614, June 1985.

[2] Cory Bennett, Ariel Tseitlin, Chaos Monkey Released Into The Wild, Netflix Tech Blog, July 30, 2012. http://techblog.netflix.com/2012/07/chaos-monkey-released-into-wild.html

[3] Jesse Robbins, Kripa Krishnan, John Allspaw, and Tom Limoncelli, Resilience Engineering: Learning to Embrace Failure, ACM Queue, Vol. 10, Iss. 9, Sept. 13, 2012. http://queue.acm.org/detail.cfm?id=2371297

[4] Inside Azure Search: Chaos Engineering, Microsoft Azure Blog, July 1, 2015. https://azure.microsoft.com/en-us/blog/inside-azure-search-chaos-engineering/

[5] Yevgeniy Sverdlik, Facebook Turned Off Entire Data Center to Test Resiliency, Data Center Knowledge, Sept. 15, 2014,

http://www.datacenterknowledge.com/archives/2014/09/15/facebook-turned-off-entire-data-center-to-test-resiliency/

[6] Kolton Andrus, Narhes Gopalani, Ben Schmaus, Failure Injection Testing, Netflix Tech Blog, Oct. 23, 2014. http://techblog.netflix.com/2014/10/fit-failure-injection-testing.html

[7] William R. Shadish, Thomas D. Cook, Donald T. Campbell, Experimental and Quasi-Experimental Designs for Generalized Causal Inference, Wadsworth Publishing, 2nd edition, January 2001.

[8] Principles of Chaos Engineering, http://www.principlesofchaos.com

[9] Philip Fisher-Ogden, Chris Sanden, & Cody Rioux, SPS : the Pulse of Netflix Streaming, Netflix Tech Blog, Feb. 2, 2015.
http://techblog.netflix.com/2015/02/sps-pulse-of-netflix-streaming.html

[10] Ding Yuan, Yu Luo, Xin Zhuang, Guilherme Renna Rodrigues, Xu Zhao, Yongle Zhang, Pranay U. Jain, and Michael Stumm: "Simple Testing Can Prevent Most Critical Failures: An Analysis of Production Failures in Distributed Data-Intensive Systems" Proceedings of the 11th USENIX Symposium on Operating Systems Design and Implementation (OSDI '14), Oct. 2014.
https://www.usenix.org/system/files/conference/osdi14/osdi14-paper-yuan.pdf

[11] Kyle Parrish, David Halsey: "Too big to test: Breaking a production brokerage platform without causing financial devastation", O'Reilly Velocity Conference, New York, NY, Oct. 2015,
http://conferences.oreilly.com/velocity/devops-web-performance-ny-2015/public/schedule/detail/45012

[12] Sam Newman: Building Microservices, O'Reilly Media, Feb. 2015.